\documentstyle[twocolumn,prb,aps,epsfig]{revtex}

\begin{document}

\title{Ocean acoustic wave propagation and ray method\\
correspondence: internal wave fine structure}
\author{Katherine C. Hegewisch, Nicholas R. Cerruti and Steven Tomsovic}
\address{Department of Physics, Washington State University, Pullman,
Washington 99164-2814}
\date{\today}
\maketitle
\begin{abstract}
\noindent Acoustic wave fields propagating long ranges through the ocean
are refracted by the inhomogeneities in the ocean's sound speed
profile.  Intuitively, for a given acoustic source frequency, the
inhomogeneities become ineffective at refracting the field  beyond a
certain fine scale determined by the acoustic wavelength.  On the other
hand, ray methods are sensitive to infinitely fine features.  Thus, it is
possible to complicate arbitrarily the ray dynamics, and yet have the
wave field propagate unchanged.  This feature raises doubts about the
ray/wave correspondence.  Given the importance of various analyses relying
on ray methods, a proper model should, at a minimum, exclude all of the
fine structure that does not significantly alter the propagated wave
field when the correspondence to the ray dynamics is integral.
We develop a simple, efficient, smoothing technique to be applied to the
inhomogeneities - a low pass filtering performed in the spatial domain -
and give a characterization of its necessary extent as a function of
acoustic source frequency.  We indicate how the smoothing improves the
ray/wave correspondence, and show that the 
so-called ``ray chaos'' problem remains above a very low frequency
($\sim 15-25$ Hz). 
\end{abstract}

\section{INTRODUCTION}

\noindent As acoustic waves propagate long ranges through the deep ocean,
they are refracted by inhomogeneities in the ocean's sound speed profile.
Roughly speaking, in the earth's mid-latitudes, temperature and pressure
effectively combine to form a wave guide in the depth coordinate that
vertically confines the propagation~\cite{munkbook}.  In addition to
this overall structure, the ocean behaves as a weakly turbulent
medium~\cite{lvov1} that multiply scatters the acoustic waves mainly in
the forward direction.  Whether one is intrinsically interested
in waves propagating through weak turbulence or in the state of the ocean
through tomography~\cite{munkbook}, ray methods are relied upon at various
stages and levels of complexity in the resulting experimental
analyses~\cite{aet1,aet2,others}.  It is therefore critical to understand
the applicability and limits of these ray methods.

Ray methods can only capture the physics of refraction and reflection,
unless a geometric theory of diffraction is explicitly
added~\cite{keller1,keller2}. Intuitively, one expects refractive effects to
dominate diffraction when sound speed inhomogeneities are larger than the
acoustic wavelength of the source.  On the other hand, due to their
pointlike nature, rays are sensitive to structures at all scales.  Thus,
one should be suspicious of (non-diffractive) ray methods for models that
have significant fine scale structure that are ineffective in refracting
waves, but that fundamentally alter the rays themselves.  Hence, certain
fine scale structures in the model can be thought of as being physically
irrelevant, i.e. having no influence on the wave propagation,
and should be eliminated before applying a ray method analysis.  The
possibility of diffraction is very important, but should be dealt with
separately and we do not discuss it further in this article.

Another serious challenge for the applicability of ray methods that has
been recognized in the past fifteen years or so, is the existence of ray
chaos~\cite{palmer}; see also earlier work in the field of quantum
chaos~\cite{berry1,berry2,zaslavsky1,zaslavsky2}.  One typical argument
goes that chaos introduces caustics, i.e.~singularities, in ray methods at
an exponentially increasing rate with propagation time (range).  Ray
methods must therefore breakdown on a logarithmically short propagation
scale, which renders them essentially useless.  A significant body of
work has shown that this need not be the case, and methods can be
developed which are accurate to much longer propagation
scales~\cite{tomsovic1,tomsovic2}. Even so, detailed ray methods tend to
become rather burdensome with the exponential proliferation of rays.
Thus, resorting to statistical methods based on the chaotic properties of
the rays is often attractive.

These two reservations about ray methods, inclusion of physically irrelevant
fine
structures in the sound speed profile and ray chaos, have often been
co-mingled.  For example, it is possible to add very fine structure to a
sound speed model that has no effect on propagating waves and yet
generates chaotic rays as unstable as one wishes.  Our point of view is
that the two issues should be disentangled, necessarily beginning with
the removal of the physically irrelevant fine structures, whose
characterization depends on the acoustic wavelength.  
We will come back to the
ray chaos question, but leave a more detailed and complete analysis
for follow-up work to this paper.

Our purpose is, thus, to create a practical and easily implemented
technique for smoothing inhomogeneities in a sound speed model, and to
give prescriptions for the extent of smoothing needed as a function of
source frequency.  Toward these ends, it is not necessary to mimic a
realistic ocean accurately with the model, but rather to include certain
key features, such as a simple form for the waveguide confinement and the
fluctuations due to the weak turbulence.  It is more than sufficient to
include scattering solely in the vertical spatial plane, to make the
parabolic approximation~\cite{tappert77} and to neglect larger
mesoscale structures.  A simple ocean sound speed model uses Munk's
canonical model~\cite{munk74} to account for large scale effects
due to temperature, pressure and salinity, and an efficient implementation
scheme by Colosi and Brown~\cite{colosibrown97} to generate
much smaller inhomogeneities due to the ocean's internal waves.
Using their approach, the inhomogeneities have the statistics of the
Garrett-Munk spectrum~\cite{gm}. 
These features, though leading to a simplified model of the ocean,
are more than adequate for investigating the length scale at which fluctuation
features become important. Increased realism will be included in a 
future companion paper~\cite{2ndpaper}.

The outline is as follows.  In Section \ref{s:sim}, 
the ocean sound speed model and
the method for acoustic propagation are presented.  The following section
considers theoretical issues such as the convergence of the propagated 
wave field by asking the question: ``does adding more modes in the
internal wave expansion cease altering the propagation beyond some
maximum mode number?''.  In Section \ref{s: smooth}, a smoothing of the expression
for the internal wave sound speed model is introduced.  This smoothing is
very similar to the application of a low-pass filter - it removes most of
the structures in the sound speed model that are shorter than a certain
scale - but it is done directly in the spatial domain so that ray
methods can easily be applied.  Sensibly, the optimal amount of smoothing
necessary is a function of source frequency.  We demonstrate the effects
of smoothing on both the wave field propagation and on the phase space
structures associated with the underlying ensemble of rays.  This does
not, in general, eliminate the consideration of ray chaos as the 
Lyapunov exponents
are still positive, but it does remove a significant amount of the
so-called ``microfolding''~\cite{simmen} of the phase space structures.
We discuss how this can markedly improve the ray/wave correspondence.

\section{THE ACOUSTIC PROPAGATION MODEL}
\label{s:sim}

In a medium such as the ocean where density fluctuations are small, the
wave equation accurately describes the acoustic waves in which we are
interested.  The governing equation is
\begin{equation}
\label{wave}
\frac{\partial^2}{\partial t^2} \Phi(\vec{r},t) =
      c^2(\vec{r},t) \nabla^2\Phi(\vec{r},t) \ ,
\end{equation}
where Re$\{\Phi(\vec{r},t)\}$ is the acoustic pressure and $c(\vec{r},t)$
is the sound speed at a location $\vec{r}$ and time $t$.  The mean sound
speed is roughly $1.5$ km/s and if we consider a water parcel, the
sound passes through it far faster than any variation in
$c(\vec{r},t)$ due to the internal waves; i.e.~the internal waves travel
several orders of magnitude more slowly than the acoustic waves.  
Hence, it is reasonable to `freeze' the state of the ocean and let
$c(\vec{r},t)=c(\vec{r})$.

In anticipation of treating long range propagation, we assume that the
scattering in the azimuthal direction is negligible and the important
components of the acoustic wave field propagation take place in two
spatial dimensions with $\vec{r}=(z,r)$, where $z$ is depth in the ocean
and $r$ is range from the source.  Consider a constant frequency source,
i.e~a pure sinusoidal source of angular frequency $\omega = 2 \pi f$ with
frequency $f$, whose
amplitude is constant in time.  Then, the wave field has a
frequency response, $\Phi_{\omega}(z,r)$, where $\Phi(z,r,t) =
\Phi_{\omega}(z,r)\, e^{- i \omega t}$.  With this assumption, the
wave equation reduces to the Helmholtz equation in cylindrical
coordinates
\begin{eqnarray}\label{eq:helmholtz}
       \nabla^2\Phi_{\omega}(z,r) + k^2(z,r) \Phi_{\omega}(z,r) = 0 \ ,
\end{eqnarray}
where the wave number $k(z,r) = \omega/c(z,r)$.

\subsection{The Parabolic Equation}
\label{ss:acoustprop}

For long range propagation, waves that propagate too steeply with
respect to the horizontal strike the ocean bottom and are strongly
attenuated. Since the surviving waves propagate at small angles with
respect to the horizontal, a Fresnel approximation~\cite{tappert77} is
possible which expresses the acoustic frequency response as the product
of an outgoing cylindrical wave, $e^{i k_0 r}/\sqrt{r}$ and a slowly
varying envelope function, $\Psi_{\omega}(z,r)$, where the horizontal
wavenumber $k_0 \approx \omega/c_0$.  Thus,
\begin{eqnarray}\label{eq:cyloutwave}
\Phi_{\omega}(z,r)&=&\Psi_{\omega}(z,r)\frac{e^{i k_0(\omega)r}}{\sqrt{r}} \ .
\end{eqnarray}
Substituting Eq.~(\ref{eq:cyloutwave}) into
Eq.~(\ref{eq:helmholtz}) and dropping two small terms gives the parabolic
equation
\begin{eqnarray}
\label{eq:parabolic}
      \frac{i}{k_0}\frac{\partial}{\partial r}\Psi_{\omega}(z,r)
= -\frac{1}{2k_0^2}\frac{\partial^2}{\partial
z^2}\Psi_{\omega}(z,r)+V(z,r)
\Psi_{\omega}(z,r) \ .
\end{eqnarray}
Since the sound speed can be decomposed into the reference sound speed,
$c_0$, and fluctuations, $\delta c$, about the reference: $c(z,r) = c_0 +
\delta c(z,r)$ with $\delta c(z,r) << c_0$, the potential is approximated
as follows:
\begin{equation}
\label{potential}
V(z,r)=\frac{1}{2}\left(1-\left(\frac{c_0}{c(z,r)}
\right)^2\right)  \approx \frac{\delta c(z,r)}{c_0} \ .
\end{equation}
In our calculations, we'll use the last form of Eq.~(\ref{potential}) for simplicity. 
Notice that there is a direct analogy between this parabolic equation and
the quantum mechanical Schr\"odinger equation through the substitutions: $t
\rightarrow r$, $m \rightarrow 1$, and $\hbar \rightarrow 1/k_0$.  We use
a symmetric split-operator, fast-Fourier-transform method to propagate the
wave field~\cite{hardintappert73,feitfleck76}; see
Appendix~\ref{ap:ssofft}.

The two terms neglected on the right side of Eq.~(\ref{eq:parabolic}) are
\begin{eqnarray}
\frac{1}{8k_0^2 r^2} \Psi_{\omega}(z,r)
+\frac{1}{2 k_0^2}\frac{\partial^2}{\partial r^2}\Psi_{\omega}(z,r) \ .
\end{eqnarray}
The first term falls off rapidly with range since $k_0 r >> 1$.  The
second term is dropped due to the parabolic approximation which assumes
that for a slowly varying envelope function, the curvature is weak.
Note that up to this point, we have also dropped other terms from the
propagation equation  in assuming negligible azimuthal scattering and
negligible time dependence of the  internal waves.  See the discussion
in Ref.~\onlinecite{flatte79} for more details on all of the terms that
have  been dropped and an order of magnitude estimate for the size of the
various contributions.

\subsection{Ocean Sound Speed Model}

A simple model for the speed of sound in the ocean consists of two main
components.  The first component of the model is an adiabatic, large
scale behavior  which is responsible for creating the ocean's `sound
channel' - an effective wave guide for acoustic propagation in the deep
ocean.  This general behavior has a minimum sound speed at the sound
channel axis, and varies slowly with latitude and season, 
with the sound channel axis moving toward the
surface for higher latitudes and colder seasons.  Mesoscale variability
is neglected in this study.  The second component
of the model is local fluctuations in the sound speed due to the ocean's
internal waves.  These fluctuations are much smaller in magnitude than
the wave guide confining behavior, but describe the range dependence.
The model potential $V(z,r)$ takes the form
\begin{equation}
V(z,r) = \frac{\delta c(z,r)}{c_0} = \frac{\delta c_{wg}(z)}{c_0}
+\frac{\delta c_{iw}(z,r)}{c_0} \ ,
\end{equation}
where $\delta c_{wg}$ represents the change of the sound speed due to the
wave guide, which we take to be range independent, and $\delta c_{iw}$
represents the fluctuations due to internal waves.

\subsubsection{The Confinement/Wave Guide}

In the ocean, the main effects of pressure, temperature, and salinity
create a minimum in the sound speed.  Since sound bends toward
regions of lower velocity, the shape of the sound speed profile refracts
propagating waves toward the sound channel axis.  This effect is captured
in a smooth, average model proposed by Walter Munk~\cite{munk74} and
is known as Munk's canonical model
\begin{equation}\label{eq:Levitus}
\frac{\delta c_{wg}}{c_0} = \frac{B\gamma}{2}
\left[e^{-\eta(z)} - 1+\eta(z)\right] \ ,
\end{equation}
where $\eta(z) =2[z-z_a]/B$,
$z_a$ is the sound channel axis, $B$ is the thermocline depth scale
giving the approximate width of the sound channel, and $\gamma$ is a
constant representing the overall strength of the confinement.  This model
has its minimum speed at $z=z_a$  and captures the right
exponential and linear trends near the surface and bottom.  The parameters
are chosen to be $B=1.0$ km, $z_a = 1.0$ km, $c_0=1.49$ km/s and
$\gamma=0.0113$ ${\rm km}^{-1}$, which are roughly consistent with
the well known environmental measurements performed in the
SLICE89 experiment~\cite{slice89-1,slice89-2}.

\subsubsection{Internal Wave Sound Speed Fluctuations}

Internal wave fluctuations perturb the sound speed in the ocean through
the resultant vertical motions of water parcels.  They are responsible for
multiple, weak, forward scattering of acoustic waves. A
numerical scheme has been introduced by Colosi and Brown~\cite{colosibrown97},
which allows efficient computation of a random ensemble of individual
realizations of the typical sound speed fluctuations.  This scheme
conforms to the Garrett-Munk spectral and statistical phenomenological
description of the internal waves~\cite{gm,viech} and has the form
\begin{eqnarray}
\label{eq:gen_iw}
\frac{\delta c_{iw}}{c_0}&=& \sum_{j=1}^{J_{max}} \sum_{k_r} e_{j,k_r}
\exp\left(-\frac{3z}{2B}\right)
\sin(j \pi \xi(z)) \ ,
\end{eqnarray}
where we took $\xi(z) = e^{-z/B}-e^{-H/B}$ with $H$ the depth of the ocean.
The prefactor $e_{j,k_r}$ includes
a random phase and magnitude factor for each $j$ and $k_r$ in the sum;
see Appendix B for further details and to infer a definition of $e_{j,k_r}$.
Since the frequency of vertical motions lie between the inertial
frequency, due to the earth's rotation, and the buoyancy frequency,
due to the local stratification,
the sum over the horizontal wave vector $k_r$ has terms representing
the superposition of internal waves with wavelengths in the range of
$1-100$ km.
A maximum for the $j$-summation has been chosen as $J_{max}=180$, 
which gives structure down to the scale of roughly a meter. 
The modes, $\sin(j \pi \xi(z))$, are connected to the buoyancy profile
which is assumed to have an exponential form. Although the form given
in the text above for $\xi(z)$
does not vanish precisely at the surface, 
it is sufficient for our purposes. 

The model should enforce that both the function $\delta c_{iw}$ 
and its derivative vanish sufficiently smoothly at the surface. 
Thus, a surface filter is introduced which
consists of multiplying Eq.~(\ref{eq:gen_iw}) by a continuous function
of depth with the properties that it vanishes above the ocean's surface,
is unity below $200$ m, and has continuous first
and second derivatives. In this way, $\delta c_{iw}/c_0$ and its derivative
vanish at the
surface and are fully, smoothly restored below $200$ m.  Since the upper
$200$ m of the ocean can be quite complex with storms, seasonal
fluctuations and latitudinal variability, there is no simple, general
sound speed model near the surface; the surface filter is
adequate for our purposes.  We will propagate waves for which very little
energy will enter this region, and thus, little effect of this surface
smoothing will be relevant.  The specific form we have
chosen for the surface filter is
\begin{equation}
\label{eq:st_filter}
g(z;z_{st}, \tau_{st}) =\left\{
       \begin{array}{ll}
        0 & \mbox{for $z' \le - 1/2 $ } \\
h(z') & \mbox{for $|z'|
\le 1/2$ } \\
1& \mbox{for $z'\ge 1/2$  }
\end{array} \right.  \ ,
\end{equation}
where $z' = (z-z_{st})/\tau_{st}$, the width is $\tau_{st} = 200$ m, 
the center is $z_{st}= \tau_{st}/2= 100$ m, and the smooth function in between
is 
\begin{equation}
h(z) = \frac{1}{2}+ \frac{9}{16}\sin(\pi z)+ \frac{1}{16}\sin(3 \pi z) \ .
\end{equation}

\subsection{Initial Wave Field}
\label{sec:wave}

The parabolic equation requires an initial wave field
$\Psi_{\omega}(z,r=0)$ as input, which can then be propagated to the
desired range of interest.  It is important to understand the connection
between the initial wave field  and the localized, continuous wave source.
Typical sources can be thought of as point sources whose acoustic energy
disperses broadly.  However, due to the previously mentioned fact that
all the steeply propagating waves are strongly attenuated, we can instead
propagate only that wave energy moving sufficiently close to the
horizontal (within a spread of angles from the sound channel axis)
that would have avoided the ocean's surface and bottom.
Restricting the propagation angles limits the size of the
vertical wave vector and necessarily creates ``uncertainty'' in the
location of the point source.
For our purposes, it is appropriate to
choose $\Psi_{\omega}(z,0)$ to be a minimum uncertainty wave packet.
This implies using a normalized Gaussian wave packet of the form
\begin{equation}
\label{ic}
\Psi_{\omega}(z,0) =\left(\frac{1}{2\pi \sigma^2_z}\right)^\frac{1}{4}
   \exp\left(
   -\frac{(z-z_0)^2}{4 \sigma^2_z} + i k_{0z}(z-z_0) \right) \nonumber  \ , \\ 
\end{equation}
where $z_0$ centers the field, $\sigma_z$ is the standard deviation of the
Gaussian intensity and $k_{0z}$ gives the propagating field an initial
wavenumber in the $z$-direction.  In all our calculations, we set
$k_{0z}=0$, which maximizes the horizontally propagating energy, and $z_0
= z_a$, which centers the energy on the sound channel axis.

A Fourier transform of Eq.~(\ref{ic}) yields a complex Gaussian
distribution of initial vertical wave numbers, $k_z$, centered at $k_{0z}$
with standard deviation in intensity, $\sigma_k$.  Since $\sigma_z^2$
and $\sigma_k^2$ are the variances of the intensity and not the amplitude
of the wave, their relation  is $\sigma_z^2 = 1/4\sigma_k^2$.   By a
simple geometrical argument, a vertical wavenumber can be related to the
horizontal wavenumber by $k_z = k_0 \tan\theta$, where $\theta$ is the
angle with respect to the sound channel axis. In the next
subsection, it is seen that $p=\tan\theta$ is a generalized momentum for
a classical ray corresponding to the wave.
Classical rays with the maximum  angle $\theta_{max}$ just
barely graze the surface or bottom, and thus, rays are limited in their
vertical wave numbers.  Yet, for Gaussian wave packets, all wave numbers
are in principle present, though most are weighted negligibly by the
tails.  It is the width, $\sigma_k$, which determines if the wave contains
wave numbers large enough for a substantial amount of the wave to hit the
surface or the ocean floor.  One can determine a proper Gaussian width,
in order for only  the Gaussian tails to pass the surface or bottom, in
analogy with the  limiting classical rays by letting the maximum
classical wavenumber $k_0\tan\theta_{max}$ correspond to three standard
deviations out in  the initial Gaussian wavenumber distribution, i.e.~set
$3\sigma_k=k_0\tan\theta_{max}$.  Then
\begin{eqnarray}
\label{eq:var}
\sigma_z^2 = \frac{9}{4 k_0^2 \tan^2\theta_{max}} \ .
\end{eqnarray}
The explicit dependence of $\sigma_z$ on the angular frequency, $\omega$,
of the continuous wave source is realized using the approximate relation
$k_0\approx\omega/c_0$.

The specific choice of $\theta_{max}$ depends on the vertical confinement.
For the background confinement in Eq.~(\ref{eq:Levitus}), those rays
departing the
sound channel axis with an angle of $\theta = \pi/18\ (10^\circ)$
come within 150 m of the surface, and those with $\theta = \pi/15\ (12^\circ)$
come within 30 m. The addition of internal waves to the sound speed model
causes some rays to have a closer approach to the surface, so
we will most often use $\theta_{max} = 10^\circ$ in this paper.

\subsection{The Classical Rays}

From the parabolic equation, one can derive a Hamiltonian system of
equations for the position, $z$, and generalized momentum, $p$,
of the collection of rays corresponding to the wave propagation.
The Hamiltonian is given by
$H= p^2/2 + V(z,r)$ and the potential is $V(z,r) = \delta c(z,r)/c_0$.
The equations are
\begin{eqnarray}
\label{dzdpdr}
\frac{dz}{dr}&=& \frac{\partial H}{\partial p} = p \nonumber \\
\frac{dp}{dr}&=& - \frac{\partial H}{\partial z}=-\frac{\partial
V(z,r)} {\partial z} \ .
\end{eqnarray}
Since $dz/dr \approx \Delta z/\Delta r = \tan\theta$,
the generalized momentum is $p=\tan\theta$.
The classical action $T$
is calculated by imposing the initial condition $T_0 = 0$ and using the
relationship
\begin{eqnarray}
\label{dTdr}
\frac{dT}{dr}&=& p \frac{dz}{dr} - H \ .
\end{eqnarray}
Through the parabolic approximation, the classical action is directly
related to the travel time, $\tau$, of the acoustic waves, where $T =
c_0\tau - r$.

The relevant rays to the wave propagation are those appropriate for a
Gaussian wave packet~\cite{tomsovic2,hhl}, which implies initial
conditions in the neighborhood of $(z_0, p_0)$.
Since $k_{0z}=k_0 p_0=0$ 
for the wave packet in Eq.~(\ref{ic}), ray calculations are done in a
neighborhood of $p_0=0$. However, $z_0$ is taken to be on the 
sound channel axis, $z_a$.

The addition of range dependent internal wave effects to the sound speed
model causes the classical rays to be chaotic~\cite{palmer}.  The
stability matrix contains the information about whether the rays are
unstable (chaotic) or not~\cite{npal1}.  At a fixed $r$,
one has
\begin{equation}
\label{vareq}
\left (
\begin{array}{c}
\delta p_r \\
\delta z_r
\end{array}
\right )
= Q_r
\left (
\begin{array}{c}
\delta p_0 \\
\delta z_0
\end{array}
\right )
\; ,
\end{equation}
where the stability matrix
\begin{equation}
\label{Q}
Q_r =
\left (
\begin{array}{cc}
q_{11} & q_{12}\\ q_{21} & q_{22}
\end{array}
\right )
= \left (
\begin{array}{cc}
\left. \frac{\partial p_r}{\partial p_0} \right |_{z_0} &
\left. \frac{\partial p_r}{\partial z_0} \right |_{p_0}\\
\left. \frac{\partial z_r}{\partial p_0} \right |_{z_0} &
\left. \frac{\partial z_r}{\partial z_0} \right |_{p_0}
\end{array}
\right )
\; .
\end{equation}
Elements of this matrix evolve according to
\begin{equation}
\label{dQdr}
\frac{d}{dr}Q_r = K_rQ_r \ ,
\end{equation}
where $Q_r$ at $r = 0$ is the identity matrix, and
\begin{equation}
\label{K}
K_r =
\left (
\begin{array}{cc}
-{\partial ^2 H\over \partial z_r \partial p_r} &
-{\partial ^2 H\over \partial z_r^2} \\
{\partial ^2 H\over \partial p_r^2} &
{\partial ^2 H\over \partial z_r \partial p_r}
\end{array}
\right ).
\end{equation}
The system of differential equations Eqs.~(\ref{dzdpdr}),
(\ref{dTdr}) and (\ref{dQdr}), are solved using a
4th order Runge-Kutta method
(where we have taken $\Delta r = 100$ m in all calculations).

The Lyapunov exponent, $\mu$, is a measure of the rate at which the
rays are deviating
under small perturbations. The relationship between the Lyapunov
exponent and the matrix
$Q_r$ comes through the trace (sum of the diagonal elements) of
$Q_r$,
\begin{equation}
\label{lyap}
\mu\equiv \lim_{r\rightarrow\infty} \frac{1}{r} \ln|Tr(Q_r)| \ .
\end{equation}
If $|Tr(Q_r)|$ grows exponentially, the Lyapunov exponent is nonvanishing
and positive, and the corresponding trajectory is chaotic.

\section{THEORETICAL CONSIDERATIONS}

Wave propagation should become increasingly insensitive to smooth
perturbations as the scale of the perturbations decreases to the order of
the smallest wavelength in the source and beyond.  
This issue does not arise in the horizontal coordinate of the 
internal wave model in Eq.~(\ref{eq:gen_iw}),
since the fluctuation scales are much longer  
than the horizontal projections of typical source wavelengths. 
However, this is an issue for the vertical fluctuations since
Eq.~(\ref{eq:gen_iw}) is a
weighted superposition of a large number of vertical internal wave
modes and presumably contains more detail than 
is necessary for
accurate wave propagation.  There comes a point in
the summation beyond which the vertical modes begin to add physically 
irrelevant features to the sound speed inhomogeneities for a given 
source frequency.  To determine the transition
point where this occurs requires an understanding of the minimum
wavelength structure in the propagating wave field, and an understanding
of the power spectrum of individual vertical internal wave modes.
The transition point, though, is not the only issue 
since higher modes contain a mix of physically relevant and irrelevant
structures. These issues as well as their interplay are discussed here.

\subsection{ The Vertical Mode Number Transition}

Intuitively, the vertical structures in the sound speed model 
responsible for refracting the wave are those that are larger than the 
minimum vertical wavelength, $\lambda_{min}$, in the initial wave packet.
Expressions for $\lambda_{min}$ can be obtained by using $\lambda = 2\pi/k$
and the previously noted geometrical relation $k_z = k_0 \tan \theta$,
\begin{equation}
\label{lmin}
\lambda_{min} = {2\pi \over k_0 \tan\theta_{max}} = 
\frac{\lambda_0}{\tan\theta_{max}}=
{c_0 \over f \tan\theta_{max}}  \ .
\end{equation}
Recall that the ocean waveguide forces $\theta_{max}$ to be small
so that the minimum vertical wavelength is always enhanced over
the source wavelength, $\lambda_0$. For $\theta_{max} = 10^\circ$, this enhancement is
roughly a factor of $6$. As a practical example, we note that
some of the experiments conducted by the
Acoustic Engineering Test (AET)~\cite{aet1,aet2} use a broadband
$75$ Hz source. A pure $75$ Hz source has a $20$ m
source wavelength. Thus, if the energy stripping due to the
ocean surface and bottom is consistent with $\theta_{max} = 10^\circ$,
then the wave propagation would have a minimum vertical wavelength scale
of roughly $110$ m.

The vertical structures in the sound speed model in Eq.~(\ref{eq:gen_iw})
arise through the superposition of vertical modes of the form
$e^{-3z/2B} \sin(j \pi (e^{-z/B}-e^{-H/B}))$.  Since the argument of the
sine is nonlinear, each vertical mode contributes different oscillation
lengths at different depths. The monotonicity of the argument illustrates
that each mode has a ``chirped'' structure, i.e.~each mode oscillates
more  and more rapidly as the surface is approached.  To make this more
precise, an expansion of the argument of the sine reveals that the local
oscillation length as a function of depth is
\begin{equation}
\label{scales}
\lambda_{iw}(z,j) = \frac{2B{\rm e}^{z/B}}{j}  \ .
\end{equation}
Therefore, the $j^{th}$ internal wave mode contributes its shortest
length contribution of $2B/j$ near the surface, with longer length scales
at increasing depth.  Each mode gives contributions to the sound speed
fluctuations over a broad range of scales.

\begin{figure}[h]
\begin{center}
\epsfig{file=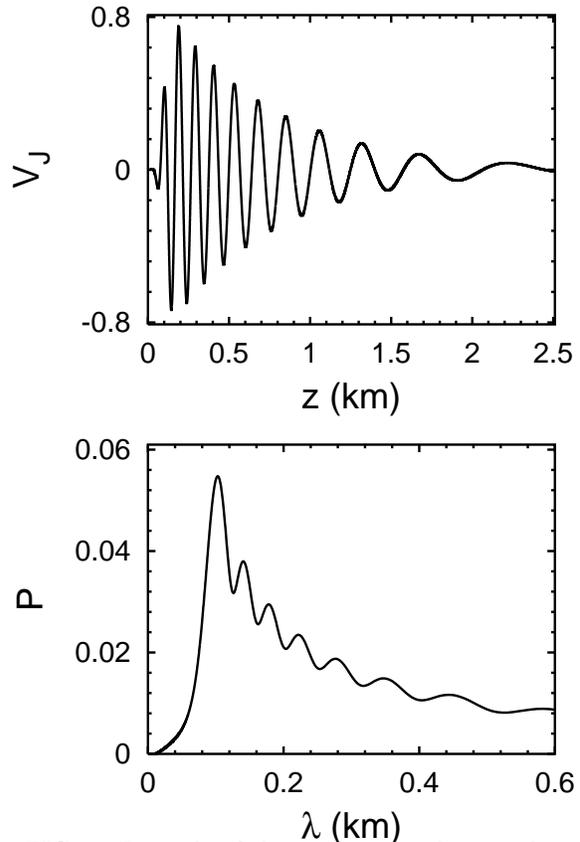, height= 4.5 in, angle =0}
\caption{Example of the single vertical internal wave mode for $j=25$.
The upper plot illustrates its depth dependence, 
$V_j=g(z;z_{st}, \tau_{st})~ e^{-3z/2B}
\sin(j \pi (e^{-z/B}-e^{-H/B}))$, 
where $g$ is the surface filter defined in Eq.~(\ref{eq:st_filter})
and the lower plot is the power spectrum, $P$, of $V_j$.}
\label{fig:jmode}
\end{center}
\end{figure}

Figure~\ref{fig:jmode} illustrates the depth dependence and power spectrum
of an internal wave mode.  The power spectrum has a fairly
sharp high frequency (short length scale) cutoff from the structures added
near the ocean surface and a slowly decaying tail for the lower
frequencies (longer length scales). The broad tail for an individual 
mode indicates that many 
different modes contribute to a
particular size feature in the internal wave model.

The transition vertical mode number $J_{trans}$ can be identified as that
point where the vertical modes begin to introduce structure smaller than
$\lambda_{min}$. Thus, setting Eqs.~(\ref{lmin}) and (\ref{scales}) equal
to each other and solving for $j$ gives
\begin{equation}
\label{minmax}
J_{trans} = \frac{2B\tan\theta_{max}}{\lambda_0} =
\frac{2Bf\tan\theta_{max}}{c_0} \ .
\end{equation}

The calculation of $J_{trans}$ does not reflect that 
each vertical mode is weighted in Eq.~(\ref{eq:gen_iw}) by the
coefficients $e_{j,k_r}$, which we numerically found to have 
root mean square decay $\sqrt{\sum_{k_r}|e_{j,k_r}|^2}\sim j^{-1.1}$
for large $j$. Thus, the higher
vertical modes have a slowly decreasing weighting. 
The acid test of
the effects of both the diminishing amplitudes and the detectability
of features by the wave is to look at the sensitivity of the wave field
to variations in the value for the $j$-summation cutoff in 
Eq.~(\ref{eq:gen_iw}). 

\subsection{ Wave Field Convergence}

We can investigate the convergence of the wave field propagation
by using different values for the $j$-summation cutoff in 
Eq.~(\ref{eq:gen_iw}) to 
generate various sound speed media.  
The value of the cutoff leading to a
converged wave field, denoted by $J_{\omega}$, is the minimum number
such that by including higher modes there is relatively little change in
the wave propagation.  We do not have a simple intuitive
argument that gives an expression for $J_{\omega}$, but instead
rely on numerical simulations to determine reasonable values.

In order to discuss quantitatively what is meant by `little change to
the wave propagation', it is necessary to have a measure of the
similarity of two wave fields.  
An ideal measure is the absolute value
squared of the overlap (inner product).
For two sound speed potentials
that differ by $\Delta V$, the quantity $C_{\Delta V}$
is defined as
\begin{equation}
\label{fidelity}
C_{\Delta V} (r) = \left|\int {\rm d}z\ \Psi^{*\Delta V}_\omega (z,r)
\Psi_\omega (z,r)\right|^2 \ ,
\end{equation}
where $\Psi_\omega(z,r)$ is understood to be the wave field propagated to
range $r$ with the full potential and $\Psi^{*\Delta
V}_\omega(z,r)$ is the same initial state propagated using the
potential which differs from the full potential by $\Delta V$.
It is convenient to normalize the propagating wave fields to unity since
this is preserved under the unitary propagation of the parabolic
equation.  With this choice, the measure gives unity only if the two
wave fields are identical.  The greater the reduction from unity, the
greater the differences between the two propagations, i.e. the lower the
faithfulness or fidelity of the propagations.

The full wave propagation is compared to wave propagation for various
values of mode number cutoff $J\le J_{max}$.  Thus, $\Delta V$ is
the internal wave sum for $j$ in the interval $[J+1,J_{max}]$.
Since deviations of $C_{\Delta V}(r)$ from unity indicate
that features in the modes $[J+1,J_{max}]$ were detectable by the
wave propagation, the value of
$J$ where $C_{\Delta V}(r)$ breaks appreciably from unity
determines $J_{\omega}$.

\begin{figure}[h]
\begin{center}
\epsfig{file=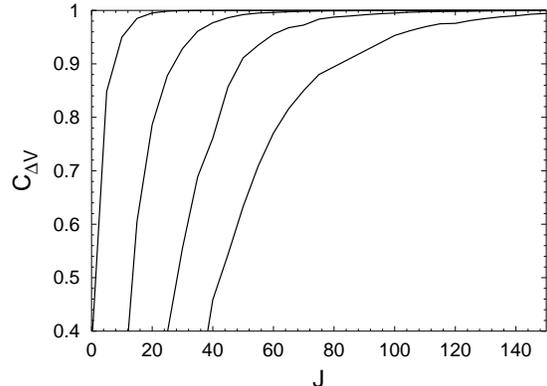, height = 3 in, angle =270}
\caption{$C_{\Delta V}(r)$ as a function of $J$
for the source frequencies of $25,75,150$ and $250$ Hz (corresponding
to the curves from left to right, respectively) at a range of 1000 km.}
\label{fig:Corrjs}
\end{center}
\end{figure}

Sound waves with source frequencies of $25, 75, 150,$ and $250$ Hz were
propagated to $r=1000$ km; see Appendix~\ref{ap:ssofft} for details
regarding the  propagation.  Figure~\ref{fig:Corrjs} demonstrates the
dependence of $C_{\Delta V}(r)$ on $J$.  To interpret this figure,
consider the curve for $75$ Hz.  Above $J=50$, 
$C_{\Delta V}(r) \ge .99$ and remains near unity.   Thus, we can say that
here $J_{\omega} \approx 50$.
Using higher internal wave modes adds no more
realism, and only slows down the calculations. A similar argument for the
other frequencies gives the values of $J_{\omega}$ listed in
Table~\ref{table1}.
Note that $J_{\omega}$ increases just a little more slowly than linear in
source frequency due, in part, to the decreasing weightings.

\begin{table}
\caption {Comparison of key parameters for a few viable
long range propagation frequencies.  Both $J_{\omega}$ and 
$\lambda_s^{opt}$ (see the next section) 
were determined
using a conservative $0.99$ criterion for the value of the
$C_{\Delta V}$ 
at $1000$ km in Figs. \ref{fig:Corrjs} and \ref{fig:Corrls}.  
Other choices for the criterion, propagation
range, etc... could lead to somewhat greater differences; however the
dependences are rather weak.  For each calculation,
$\theta_{max} = 10^\circ$.
Note the minimum wavelength
feature, $\lambda_{min}$, in the initial wave packet is extremely close
to $\lambda_s^{opt}$.}
\begin{tabular}{cccccc}
Frequency  & $J_{trans}$ & $J_{\omega}$ & $\lambda_0$  &
$\lambda_{min}$  & $\lambda_s^{opt}$  \\
(Hz) & & & (km)& (km) & (km) \\ \hline
$25$ & 6 & 20 & 0.060 & 0.340 & 0.308 \\
$75$ & 18 & 50 & 0.020 & 0.113 & 0.106 \\
$150$ & 36 & 90 & 0.010 & 0.056 & 0.060 \\
$250$ & 60 & 145 & 0.006 & 0.034 & 0.034 \\
\end{tabular}
\label{table1}
\end{table}

\begin{figure}[h]
\begin{center}
\epsfig{file=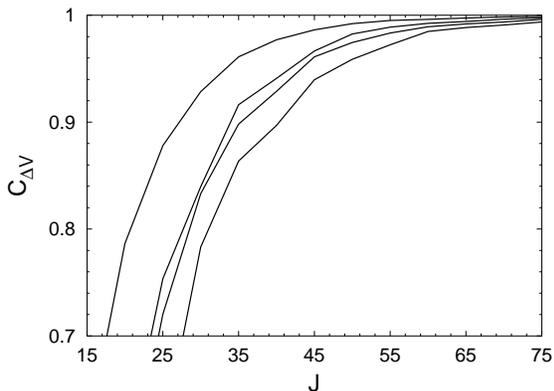, height = 3 in, angle =270}
\caption{$C_{\Delta V}(r)$ as a function of $J$
for the ranges of $1000,2000,3000$ and $4000$ km (corresponding
to the curves from left to right, respectively) for a frequency of
75 Hz.}
\label{fig:Corrjrange}
\end{center}
\end{figure}

Since $C_{\Delta V}$ is inherently range dependent,
determining $J_{\omega}$ from a plot of $C_{\Delta V}$ for a single range
is potentially inappropriate for longer ranges. Yet,
Figure~\ref{fig:Corrjrange} illustrates that the range
dependence of $C_{\Delta V}$ is fairly weak for a frequency of
$75$ Hz. Increasing the range from $1000$ km to $4000$ km for
$J_{\omega}=50$ decreases $C_{\Delta V}$ very little
from $0.99$ to $0.96$.
Thus, $J_{\omega}=50$ is a conservative choice
even for ranges up to $4000$ km.

For reasonable source frequencies, it is clear that even an optimal
choice for $J_{\omega}$ will leave a significant amount of oscillations in
the model on a scale much smaller than $\lambda_{min}$ (since
$J_{\omega} >> J_{trans}$).  Presumably, these oscillations are
physically irrelevant for the wave propagation, but it is worthwhile
studying more precisely where the cutoff lies within the context of
long range propagation.

\section{FILTERING THE PHYSICALLY IRRELEVANT FEATURES}
\label{s: smooth}

Since we have taken a smooth background sound speed model,
the physically irrelevant features of the sound speed model
can be removed by filtering the high
frequency components from the internal wave sound speed model, $\delta
c_{iw} (z,r)$. The ideal approach would be through the application of a
low pass filter:  Fourier transform $\delta c_{iw} (z,r)$ for a fixed
range to a frequency domain, apply a filter that removes the high
frequencies and Fourier transform back to give the physically relevant
portion of $\delta c_{iw} (z,r)$.  There are several drawbacks with
respect to proceeding this way.  The Fourier transforming back and forth
is computationally expensive, creates a problematic ocean surface, and
severely complicates the ray correspondence; the same would be true using
a convolution technique.  Instead, we develop a smoothing that can be
directly applied to each vertical mode in the spatial $z$ domain and
serves as a very good approximation to a  low-pass filtering in the
frequency domain.  It takes advantage of the monotonicity of the chirped
structure of the individual internal wave modes.   The spatial filtering
method simplifies the ray equations enormously and allows first and second
derivatives to be evaluated exactly, as opposed to numerically, which is
an unstable operation.

\subsection{The Smoothing}

Due to the precise oscillatory nature of each vertical mode, a
good approximation to a low-pass filter can be accomplished by removing
the upper portion of the vertical mode that contains oscillations
that are smaller than the smoothing parameter, $\lambda_s$.
This involves multiplying
each vertical mode by the function $g(z;z_{sm},\tau_{sm})$
defined in Eq.~(\ref{eq:st_filter}).
This filter is centered at the depth such that the local
length scale is $\lambda_s$, which by inversion of Eq.~(\ref{scales}),
gives the mode-dependent depth $z_{sm}=B\ln(j\lambda_s/2B)$.  Note that
$j$ must exceed $2B/\lambda_s$ in order for the filter to be below the
ocean surface, which is where it begins to have an effect.
This is consistent with the shortest length contribution of each
mode being $2B/j$ at the surface.
The width of the filter is carefully chosen to be $\tau_{sm} =
2.0\lambda_s$ so that it does not cut off too sharply thereby
introducing high frequency components into the model. If the width
were chosen much greater, amplitudes of physically relevant length scales
would be reduced.
\begin{figure}[h]
\begin{center}
\epsfig{file=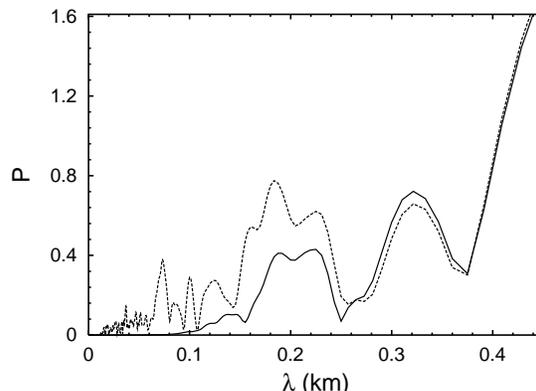, height = 3 in, angle =270}
\caption{Effects of the amount of smoothing on the power spectrum, $P$,
of $\delta c_{iw}/c_0$.
The dashed line is the power spectrum of the unsmoothed full potential and
the solid line is the power spectrum of the smoothed full potential
for $\lambda_s=0.20$ km.}
\label{fig:smoothinglambda20}
\end{center}
\end{figure}

\begin{figure}[t]
\begin{center}
\epsfig{file=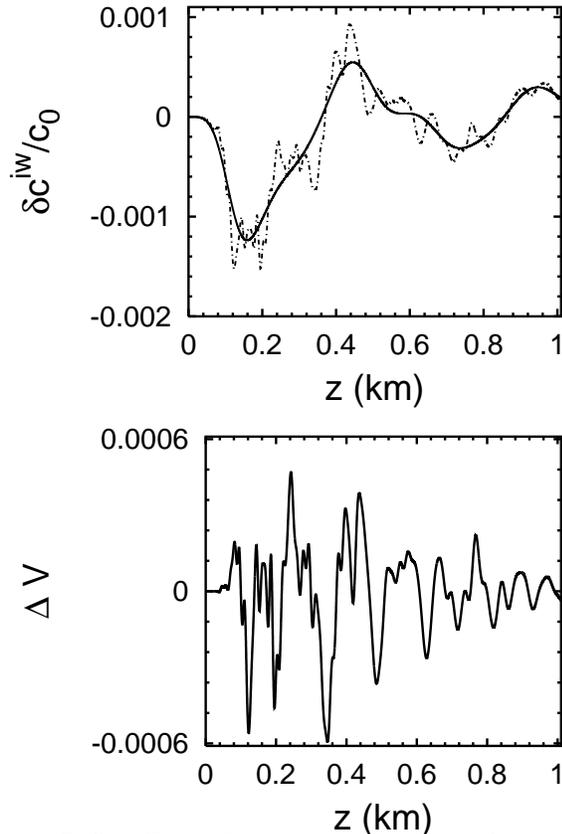, height = 4.5 in, angle =0}
\caption{Effects of the amount of smoothing on the full potential, $\delta
c_{iw}/c_0$.
In the upper panel, the dashed line is the unsmoothed potential and
the solid line is the smoothed potential for $\lambda_s=0.20$ km.  In the
lower panel, the difference, $\Delta V$, between the smoothed and unsmoothed
potential is displayed.  }
\label{fig:smoothings}
\end{center}
\end{figure}

Figure \ref{fig:smoothinglambda20} shows the power spectrum of the sound
speed model with and without smoothing; it is illustrated with a value,
$\lambda_s=0.2$ km.  The power spectrum remains relatively unchanged
for length scales greater than $0.2$ km, but the length scales below $0.2$ km
are significantly dampened out of the model.   This is evidence that a
smoothing parameter of $\lambda_s=0.2$ km is doing exactly what it was
designed to do: it is filtering out features on scales below $0.2$ km, but
leaving features above $0.2$ km in the model.  Figure~\ref{fig:smoothings}
shows the smoothed sound speed potential and the portion of the potential,
$\Delta V$, filtered by the smoothing.  It is clear from these figures
that the oscillations in the unsmoothed potential which have a
length scale of less than $0.2$ km have been removed, while larger
oscillations have been preserved.

\subsection{Estimating the Optimal Smoothing Parameter}

The optimal smoothing parameter, $\lambda_s^{opt}$, would be such that only
those features in the model that are not detectable by the wave would be
removed.  Intuitively, $\lambda_s^{opt}$ would be very close to
$\lambda_{min}$ of Eq.~(\ref{lmin}).  In order to test this intuition,
we again use $C_{\Delta V}(r)$ defined in
Eq.~(\ref{fidelity}), where here $\Delta V$ is the high frequency portion
of the internal wave sum, which the smoothing removes, and the other potential
is the full unsmoothed sound speed model. $\lambda_s^{opt}$ is
determined to be the maximum  value of $\lambda_s$ up to which $C_{\Delta
V}$ remains nearly unity  but deviates significantly beyond.
\begin{figure}[h]
\begin{center}
\epsfig{file=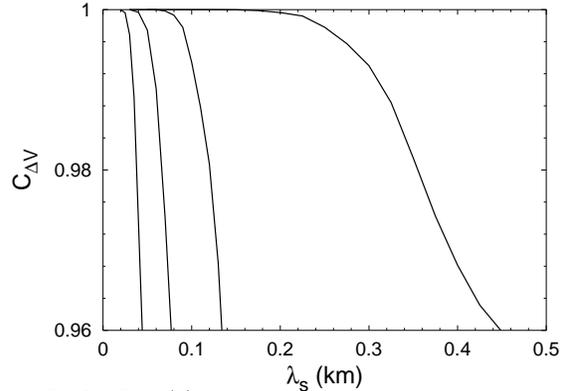, height = 3 in, angle =270}
\caption{$C_{\Delta V}(r)$ as a function of $\lambda_s$
for source frequencies of $25,75,150$ and $250$ Hz (corresponding
to the curves from right to left respectively) at a range of 1000 km.}
\label{fig:Corrls}
\end{center}
\end{figure}

As in the previous section, source frequencies of $25, 75, 150,$ and $250$ Hz
were propagated to $r=1000$ km with $J$ chosen for each
frequency to be that value of $J_{\omega}$ in Table~\ref{table1}.
Figure~\ref{fig:Corrls} demonstrates the
dependence of $C_{\Delta V}(r)$ on different values of $\lambda_s$
and its interpretation is similar to that done for Fig.~\ref{fig:Corrjs}.
Consider the curve for $75$ Hz.  Above $\lambda_s \approx 0.1$ km,
$C_{\Delta V}$
breaks significantly from unity giving the optimal smoothing of
the sound speed model for a $75$ Hz source to be $\lambda_s^{opt} \approx 0.1$
km.  Smoothing less than this allows high frequency features to remain in
the model which have no effect on the wave propagation.
Table~\ref{table1} summarizes the results which all agree closely with
the intuitive idea that $\lambda_s^{opt}\approx \lambda_{min} =
\lambda_0/ \tan \theta_{max}$. 

For a fixed $\lambda_s$, the higher source frequencies lead to a reduced
value of $C_{\Delta V}$.
This indicates that  the high frequency components of
$\Delta V$ are more detectable by a high frequency source than by a low
frequency source.  This fully supports the age-old intuitive concept that
high frequency waves can detect smaller features than low frequency waves,
and that the appropriate detection scale is a wavelength.  A long
range propagation experiment utilizing a source frequency $f$ only detects
that portion of the internal wave power spectrum with features longer than
the length scale $\lambda_{min}=c_0/ f \tan \theta_{max}$.

\subsection{Effects of Smoothing on Phase Space Structures}
\label{ss: phasespace}

Classical ray methods can be used to reconstruct propagating wave fields
in detail through the use of semiclassical Green
functions~\cite{schulman}. The semiclassical approximation to the
wave field is
\begin{equation}
      \label{semi}
      \Psi_{sc}(z, r; k_0) = \sum_j A_j(z,r) \exp[i k_0 T_j(z, r) - i\pi \nu_j /2] \ ,
\end{equation}
where the sum is over all ray paths labeled by $j$
that begin at the source and end at a depth $z$ for a given range $r$.
The phase contribution of a path is related to  
its classical action, $T_j$, the source wavenumber, $k_0$,
and the number of caustics, $\nu_j$.   
The amplitude contribution of a path, $A_j$, is related to its
stability matrix elements; see Ref.~\onlinecite{heller} for a readable account.
This discrete set of paths becomes continuous if we consider all $z$.  Thus,
there is a continuous set of rays that underlies the full construction of
$\Psi_{sc}(z, r; k_0)$ at a given range.  A powerful analysis of the
properties of this set comes by considering the rays in the phase space
formed by all allowable points given by position and conjugate momentum.
Viewed in phase space, the continuous set of rays underlying the wave field
propagation (in the single degree of freedom problem being discussed
here) forms a continuous, self-avoiding line which is called a
manifold.  As the range increases, the manifold evolves into a rather
wild-looking ``spaghetti''.  The more chaotic the system, the wilder the
appearance of the manifold.

The construction of Eq.~(\ref{semi}) relies on the use of stationary phase
approximations, which can only be applied reliably when the phase between
successive stationary phase points is greater than order unity.  Care
must be taken in defining the meaning of successive in this context.
Thus, Eq.~(\ref{semi}) breaks down when $[T_j(z, r) - T_{j'}(z, r)] <
k_{0}^{-1} = \lambda_{0}/2\pi$ where $j$ and $j'$ are the classical
paths/rays corresponding to successive stationary phase points.  
We term this the
`area-($\lambda_{0}/2\pi$) rule' (the translation to this
problem of the area-$\hbar$ rule of Refs.~\onlinecite{berry1} and
\onlinecite{berry2}).  See Refs.~\onlinecite{tomsovic1} and
\onlinecite{tomsovic2} for a detailed presentation of the breakdown of
the stationary phase approximation in quantum chaotic systems.

The breakdown of stationary phase is intimately related to how the
manifold winds and folds its way through phase space.  The difference in
the classical action for two rays is related to the areas in phase space
between the folds of the evolving manifold and the vertical line of the
final depth, $z$, whose intersections with the manifold specify the
rays.  If these areas become smaller than $\lambda_{0}/2\pi$, then
stationary phase breaks down for that pair of rays and we say that the
two stationary phase points are coalescing.  By drawing the manifold and
filling in areas of $\lambda_{0}/2\pi$ in the folds, one can immediately
see where problems, such as caustics which produce infinite amplitudes,
will be occurring in the
semiclassical construction.  In the simplest case of two coalescing
points, an Airy function uniformization is possible.  However, if so many
coalescing pairs occur that they cannot be isolated from each other,
uniformization effectively is no longer possible, and the semiclassical
approximation has broken down.

In the work of Simmen, Flatt\'{e}, and Wang~\cite{simmen}, they show how the
fine features in the internal wave field lead to a phenomenon they termed
``micro-folding'' in which tiny folds are densely found along the
manifold.  Clearly, for typical source frequencies in long range
propagation, the neighborhoods of the micro-folds violating the
area-($\lambda_{0}/2\pi$) rule overlap everywhere with each other.  Thus,
one anticipates a dense set of singularities in the semiclassical
approximation rendering the approach useless.
\begin{figure}[h]
\begin{center}
\epsfig{file=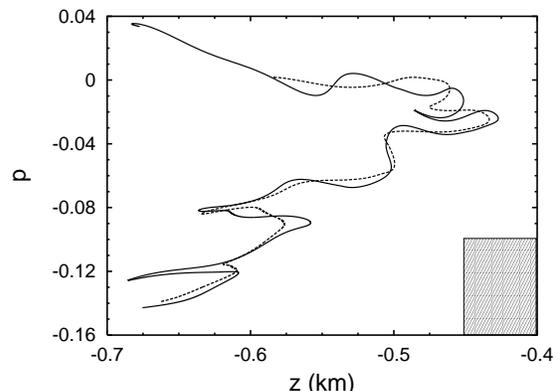, height = 3 in, angle =270}
\caption{Smoothed phase space manifold.  The solid line is the phase space
plot for a dense set of trajectories with launch angle
$\theta \in [4^\circ, 8^\circ]$ propagated for $50$ km in the unsmoothed
ocean model.  The dashed line is for the same set of trajectories, but
for a smoothing parameter of $\lambda_s = 0.10$ km.
All the trajectories started on the sound channel axis.
The hatched rectangle is a
reference area for physically irrelevant microfolds and has an area 
$\lambda_0/2
\pi$, which corresponds to a $75$ Hz source.}
\label{fig:phasespace_big}
\end{center}
\end{figure}

\begin{figure}[h]
\begin{center}
\epsfig{file=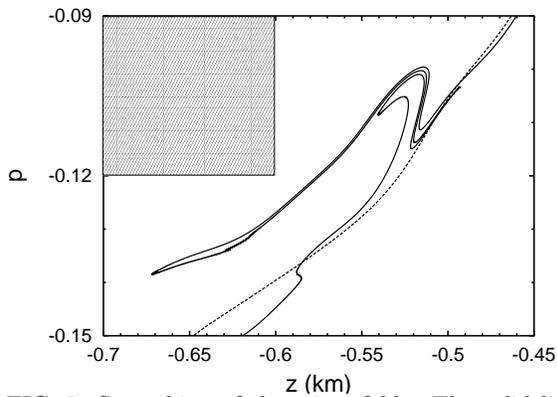, height = 3 in, angle =270}
\caption{Smoothing of the micro-folds.  The solid line is the phase space
plot for a dense set of trajectories with launch angle
$\theta \in [7^\circ, 8^\circ]$ propagated for $100$ km in the unsmoothed
ocean model.  The dashed line is for the same set of trajectories, but
for a smoothing parameter of $\lambda_s = 0.10$ km.
All the trajectories started on the sound channel axis.
The hatched rectangle is a
reference area for physically irrelevant microfolds and has an area 
$\lambda_0/2 \pi$,
corresponding to a $75$ Hz source. }
\label{fig:phasespace_zoom}
\end{center}
\end{figure}

Herein lies the advantage of smoothing the ocean sound speed model of
physically irrelevant features before making the ray correspondence.  
Presumably,
the bulk of the micro-folding is related to fine features which are
ignored by the wave propagation.  The smoothed system contains only that
structure necessary to describe the wave propagation so it should have
fewer micro-folds.  Figure~\ref{fig:phasespace_big} illustrates the
effects of smoothing on a set of trajectories.  One can see that the
smoothed manifold tracks the  unsmoothed manifold along its length very
well.  A more detailed example of micro-folding is illustrated in
Fig.~\ref{fig:phasespace_zoom} for a range of $100$ km.  Notice how the
smoothed manifold completely eliminates this particular micro-folded
structure for a smoothing parameter of $\lambda_s = 0.1$ km (appropriate
for $75$ Hz).  Eleven, non-isolated pairs of coalescing stationary phase
points were eliminated by the smoothing.  Only a well behaved piece of
the manifold with no coalescing pairs remains.  Thus, there are fewer
locations leading to singularities and breakdown in the semiclassical
construction for the smoothed system, yet it is describing the same
propagated wave.  We leave the full semiclassical reconstruction for
future work.

\subsection{Effects of Smoothing on Lyapunov Exponent}

\begin{figure}[h]
\begin{center}
\epsfig{file=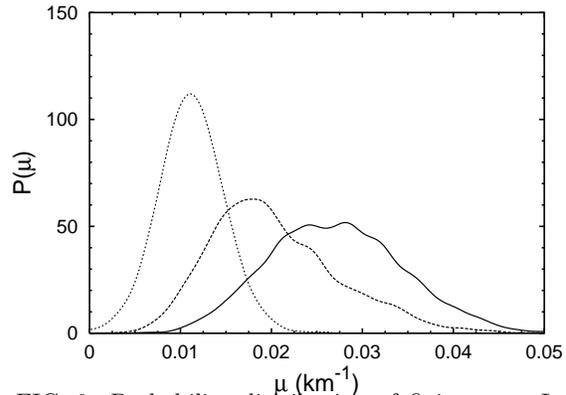, height = 3 in, angle =270}
\caption{Probability distribution of finite-range Lyapunov exponents.
The range of propagation is $1000$ km and each probability distribution
consists of 4,000 trajectories within a uniform distribution of launch angles.
For the solid line, $|\theta| \in [0^\circ, 2^\circ]$, for the dashed
line, $|\theta| \in [4^\circ, 6^\circ]$, and for the dotted line,
$|\theta| \in [8^\circ, 10^\circ]$.  Each probability distribution was
obtained by averaging over a Gaussian window of the corresponding histogram.
The smoothing parameter is $\lambda_s = 0.10$ km
and all the trajectories started on the sound channel axis.}
\label{fig:lyap-hist}
\end{center}
\end{figure}

The following question naturally poses itself from the results of the
previous section, ``if smoothing the inhomogeneities reduces the number
of folds, perhaps it is eliminating the ray chaos that was discovered
in Ref.~\onlinecite{palmer}?''  This turns out not to be the case.  The
Lyapunov exponents for smoothed systems do not vanish.  The Lyapunov
exponent, $\mu$, as defined in Eq.~(\ref{lyap}), requires the infinite
range limit, which due to the maximum range of the ocean, is not very
sensible.  Instead, it is much more relevant to work with finite-range
Lyapunov exponents~\cite{grass,wolfsontomsovic}.  The stability matrix,
$Q_r$, as defined in Eq.~(\ref{Q}), is calculated for a classical ray
starting on the sound channel axis with an initial angle $\theta$ and
propagated for a range $r$. If $|Tr~Q_r|$ is growing exponentially with
range, then the ray is unstable or chaotic and the following relationship
can be inverted to obtain the finite-range Lyapunov exponent
\begin{equation}
|Tr~Q_r| = e^{\mu r}+e^{-\mu r}  \ .
\end{equation}
Excluding a few highly abstract systems, this $\mu$ fluctuates as a
function of range and from one ray to the next.  In fact, for typical
chaotic systems and the internal wave problem here, $|Tr~Q_r|$ is close to
being lognormally distributed, or from a different point of view, the
finite-range Lyapunov exponents give something close to a Gaussian
density~\cite{grass,wolfsontomsovic}.  The finite-range Lyapunov exponents
are launch angle dependent~\cite{vera}.  Figure~\ref{fig:lyap-hist}
shows histograms of the finite-range Lyapunov exponents for a range of
$1000$ km for a range of ray angles.

\begin{figure}[h]
\begin{center}
\epsfig{file=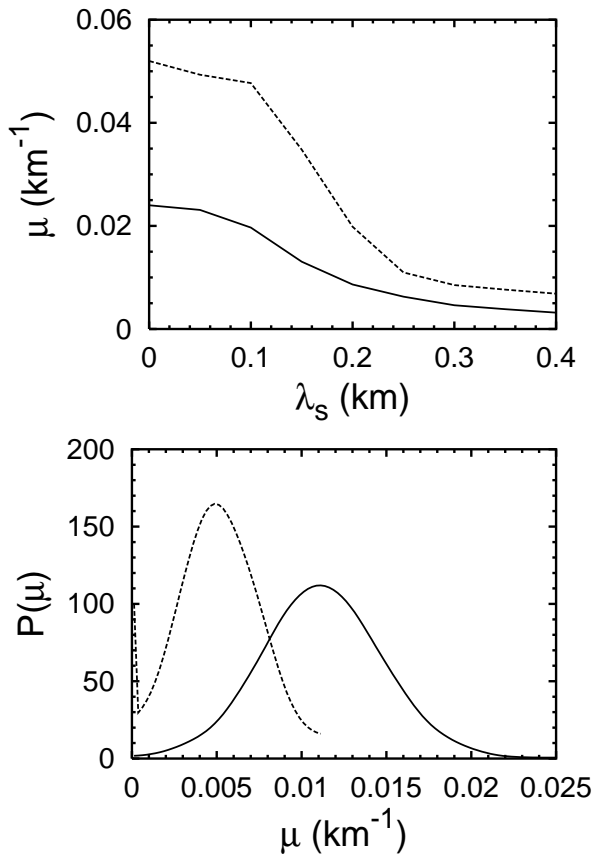, height = 4.5 in, angle =0}
\caption{Average Lyapunov exponents, $\mu_0$ and $\bar{\mu}$,
and probability distribution as a function of the smoothing parameter,
$\lambda_s$.  Both plots are for a range of propagation of $1000$ km.
The upper plot is an average of 2,000 trajectories within a uniform
distribution of launch angle $\theta \in [-10^\circ, 10^\circ]$.  The
solid line is $\mu_0$ and the dashed line is $\bar{\mu}$.  The lower plot
is the same as the previous plot except that $|\theta| \in [8^\circ,
10^\circ]$ in both curves and the smoothing parameter is varied.  The
solid line is for a smoothing parameter of $\lambda_s = 0.10$ km and the
dashed line is for a smoothing parameter of $\lambda_s = 0.30$ km.  A
narrow peak near the origin exists in the dashed curve, which indicates a
non-negligible fraction of stable trajectories.}
\label{fig:lyap-smooth}
\end{center}
\end{figure}

It turns out that the mean of the finite-range Lyapunov exponents is the
usual infinite-limit Lyapunov exponent (as long as one has propagated
beyond a transient range of a few Lyapunov lengths).  Letting the
brackets $<>$ denote averaging over many rays,
\begin{equation}
      \mu_0 =\frac{1}{r}<\ln{|Tr~Q_r|}> \ .
\end{equation}
If one averages before taking the natural logarithm, one gets a second
stability exponent which is not the Lyapunov exponent, but rather a
related one:
\begin{equation}
      \bar{\mu} = \frac{1}{2r} \ln(<|Tr~Q_r|^2>) \ .
\end{equation}
The relationship between $\mu_0$ and $\bar{\mu}$ for a Gaussian density is
through the variance of the distribution of the finite-range Lyapunov
exponents
\begin{equation}
      \sigma^2_\mu = \frac{\bar{\mu} - \mu_0}{r} \ .
\end{equation}
These two stability exponents fix the Gaussian density completely.
Figure~\ref{fig:lyap-smooth} illustrates the dependence of $\mu_0$,
$\bar{\mu}$ and the distribution on the smoothing parameter $\lambda_s$.

Although, there is still ray chaos, the Lyapunov exponent is 
monotonically decreasing with increased smoothing, but 
unless a smoothing greater than $0.10$ km is applied, $\mu_0$ 
does not decrease appreciably.  At some point, beyond
a smoothing somewhere in the neighborhood of $0.3-0.5$ km, a large
fraction of the rays behave stably.  Note in Fig.~\ref{fig:lyap-smooth}
that for $\lambda_s = 0.30$ km, a significant fraction of the rays
have become stable, i.e., they have a Lyapunov exponent equal to zero.
Using the relation between frequency
and optimal smoothing, for source frequencies in the neighborhood of
$15-25$ Hz, there is a transition below which
the ray chaos problem due to the internal wave inhomogeneities
effectively disappears and above which it remains important over ocean basin
scale propagation ranges.  
Though the background profile used for this study is 
somewhat simplistic, surprisingly 
these results seem to be consistent with some
very low frequency experiments. In particular, data from the Alternate
Source Test (AST) clearly shows that $28$ Hz receptions have a
more stable arrival pattern than the $84$ Hz receptions for transmission over
a $5000$ km range~\cite{ast}.

\section{DISCUSSION}

In probing the state of the ocean, it is
important to understand what information is carried in the wave
propagation for a given source frequency.
Intuitively, fluctuations in
the ocean sound speed on
scales shorter than an acoustic wavelength should be ineffective sources
of refraction for a sound wave in the ocean.
Though, parabolic equation simulations are unaffected by
the inclusion of physically irrelevant fine scale fluctuations
in the sound
speed model (except for the resulting slower computation time), this
inclusion worsens the correspondence of ray methods to the wave
propagation.  
On the other hand, ray methods are sensitive to infinitely fine scale structures.  Those fine
structures that are not detectable by the wave propagation lead to
physically irrelevant micro-folds in the phase space manifolds for the rays.
These
folds lead to unwanted singularities and the breakdown of semiclassical
approximations.  Smoothing of the internal wave sound speed model allows
a significant reduction in the extent of micro-folding and this must lead
to a better ray/wave correspondence.

In our study, we noted
that the chirped structure of each of the internal wave modes gives
contributions to the sound speed fluctuations over a broad range of
scales.  Thus, limiting the number of vertical modes used in an internal
wave sound speed model does not fully resolve the issue of physically
irrelevant
fine structure.  For the specific construction of Colosi and Brown, our
calculations gave frequency dependent values for the number of
vertical modes $J_{\omega}$ necessary in the model.  For frequencies of
$\{25,75,150,250\}$ Hz, we found that the wave field propagation
is essentially converged
for $J_{\omega}=\{20,50,90,145\}$,
respectively.  However, for the same set of frequencies, modes greater
than the transition modes $J_{trans}=\{6,18,36,60\}$, respectively,
add structures on a finer scale than
$\lambda_{min}$.  Hence, each mode contains a large spread of frequency
contributions so that a low-pass filtering of each vertical mode is needed.

In order to remove physically irrelevant structures,
we constructed an approximation in the position domain to a low-pass filter 
by taking advantage of the monotonicity of the chirped structure of
each mode.  The accuracy of this approximation (though not shown in this
paper) was very good for individual modes.  The spatial filtering method
that we developed
gives three main advantages: reducing required computations, better behavior
in the neighborhood of the ocean's surface, and simplicity with respect to
making the ray correspondence. With this study, it was 
found that the vertical scale of interest for the vertical fluctuations
is not the source wavelength, $\lambda_0=c_0/f$,
but rather the minimum  vertical wavelength present in the wave field,
which contains the additional projection factor $(\tan\theta_{max})^{-1}$; see
Eq.~(\ref{lmin}).  $\theta_{max}$ is the largest angle with respect to the
horizontal that waves can propagate without being stripped out by bottom
interactions and is typically in the neighborhood of $10^\circ-12^\circ$
in the ocean's mid-latitudes. For these values of $\theta_{max}$ the
minimum vertical wavelength is roughly $5-6$ times $\lambda_0$;
i.e. relevant vertical structures are much larger than that
implied by $\lambda_0$.

Additionally, from the results in
Table~\ref{table1}, $J_{\omega}$ scales more slowly with
increasing frequency than $J_{trans}$.
This appears to be due to the decreasing weighting of the terms in
Eq.~(\ref{eq:gen_iw}), which directly influences the convergence
of the wave field propagation and the value of $J_{\omega}$.
If this trend were to continue, then at a sufficiently high frequency,
$J_{trans}$ would overtake $J_{\omega}$ in value. 
Beyond this frequency, low-pass filtering would no longer
serve any purpose; one could simply choose an appropriate $J_{\omega}$.
We do not attempt to extrapolate to that point here using our calculations
and model,
but note that wherever it is, the frequency would be so high that very
long-range acoustic propagation would not be possible due to dissipation. 
However, it may be useful in the context of short range acoustic
experiments using much higher frequencies to establish a cross-over frequency
with a more realistic model.

We found that smoothing the internal wave sound speed fluctuations
does not,
in general, eliminate the problems associated with ray chaos.  The
Lyapunov exponents are positive and significant unless the smoothing
scale exceeds 300 - 500 m.  Thus, in this simplified model, ray chaos
continues to be an important issue for source frequencies above the
15 - 25 Hz range. 

A number of difficulties arise in the study of chaotic systems.  
For example, the exponential prolification of rays, makes it impractical
to carry out ray methods.  A common technique to overcome these 
difficulties is to apply various statistical methods whose justification
derives from the chaos itself. However, even if you wish to apply these
statistical methods, the validity of semiclassics is still an issue. 

Though it is known in the literature that the background sound speed profile
can dramatically affect the complexity of the ray dynamics, it is still a
question for investigation as to how significant these effects are on the
wave propagation.
Here we use Munk's canonical model as a simple, smooth
background profile,
which is sufficient for a study of the
removal of physically irrelevant structures.
However, before inferring detailed properties of long range
experimental data, it would be good to employ a more realistic background
sound speed wave guide.
In fact, this would require a method for removing fine scale structures
from the background in addition to the internal wave model and
would not likely be subject to as
simple a spatial filtering scheme as we used for the internal waves.
We will address these issue in a forthcoming paper.

A number of interesting questions remain or emerge from our results.
Our computations did not use pulsed sources,  which can be expressed as
an integral over a range of frequencies.   It would seem reasonable to
assume that the dynamics should be smoothed less for higher frequencies
than for appreciably lower frequencies.  How much attention must be
paid to this issue?  Can one make the crude approximation of using
smoothing for the center frequency of a pulse?

In pulsed experiments, the early arrivals form branches which correspond to
wave energy propagating at the larger angles near $\theta_{max}$.
Depending on the range of propagation, the late arrival portion of
the signal may
be constrained to a narrower range of propagation angles.  Is more
smoothing appropriate for this portion due to the $\theta_{max}$ factor
in $\lambda_{min}$?

The measure $C_{\Delta V}$  
is quite generally a function of range.  Yet, we mainly used $1000$
km propagation in our calculations to determine the optimal amount of
smoothing and neglected the range dependence; we did note however a weak
range dependence.  
Recall that several approximations are made arriving at the
parabolic equation or other one-way, small-angle approximation variants.
The neglected terms may also put range dependence in the propagation, and
it would not make sense to try to be more accurate with the smoothing
than the level of these other approximations.  A more detailed
understanding of the effects of neglected terms would be desirable.

Although, there is significantly less micro-folding for the smoothed
than for the unsmoothed potentials, there is still uncertainty as to how
much improvement is gained for the optimal smoothing.  This could be made
clear by carrying out the full detailed semiclassical construction and
comparing it to the wave field propagation; we will carry this out in 
Ref.~\onlinecite{2ndpaper}. 
A much deeper understanding would come from a full
theory based on applying the area-($\lambda_{0}/2\pi$) 
rule discussed in Sec.~\ref{ss: phasespace}.
It would give the most precise answers possible to questions of which structures 
are physically irrelevant and which method removes them in the most optimal
way.  We are pursuing this investigation 
because only by separating out the physically irrelevant fine
structures can we begin to fully address the ray chaos conundrum and
know whether it can be overcome.

\section*{ACKNOWLEDGMENTS}

We gratefully acknowledge M.~G.~Brown for a critical reading of the
manuscript and for the support of the Office of Naval Research
through the grant N00014-98-1-0079 and the National Science Foundation
through the grant PHY-0098027.

\appendix

\section{THE SPLIT-OPERATOR, FAST FOURIER TRANSFORM METHOD}
\label{ap:ssofft}

The parabolic equation in Eq.~(\ref{eq:parabolic}) describes the
propagation of an acoustic wave  with Hamiltonian $H = p^2/2 + V$, where
$p^2/2$ and $V$ denote the  kinetic and potential energies. A wave field
can be advanced exactly through the
application of the unitary propagation operator $\exp(-i k_0 \int H dr)$.
The split-operator Fourier transform method~\cite{feitfleck76}
approximates this operator using
$e^{A+B}\approx e^{A/2} e^B e^{A/2}$,
where  $A$ is taken to be $-i k_0 \int (p^2/2)~dr
= -(i/k_0) \int (k^2/2)~dr $ and $B$ is taken
to be $-i k_0 \int V(z,r) dr$. Inserting a Fourier transform identity and
rearranging terms before integrating gives a formula for the propagation
   of a wave field, $\Psi_\omega(z,r)$, at a range $r$ to a wave field, 
$\Psi_\omega(z,r')$,
   at a range $r'=r + \Delta r$
   \begin{eqnarray}
       \label{splitstep}
       \Psi_\omega(z,r') &=&F^{-1}\left[ e^{A/2}
               F\left[ e^{B}
               F^{-1} \left[ e^{A/2}
               F\left[ \Psi_\omega(z,r) \right] \right]\right]\right] 
\nonumber \\
   \end{eqnarray}
   where $F$ and $F^{-1}$ are the forward and backward Fourier transforms,
   respectively.  This equation has error
$O\left(\left( k_0~\Delta r \right)^3\right)$ due to the operator
approximation.
We approximate the integral $\int_{r}^{r'} V(z,r)dr \approx \Delta
r~[V(z,r)+
V(z,r')]/2$ and perform the integration
$\int_r^{r'} (k^2/4)~dr = \Delta r ~ k^2/4 $.

The wave fields in this paper are calculated over a vertical grid of
$z\in[-2,7]$ km.
The reflection boundary condition at the surface is not enforced
in favor of the wave experiencing a soft reflection from the
potential rather than a hard reflection from the surface.
Wave energy which is reflected from the surface
is eventually absorbed by the bottom in long range propagations
so that this energy is negligible
at a receiver.
The soft reflections of the wave are due only to the background portion of the
potential (Munk's canonical model in Eq.~(\ref{eq:Levitus})) whose
effects have been
extended above the surface, $z<0$. The internal wave fluctuations from
Eq.~(\ref{eq:gen_iw}) are cut off by the surface filter in
Eq.~(\ref{eq:st_filter}) so that they don't
have an effect on the wave above the surface.

The grid size for the propagation is chosen to be dependent on the
source frequency (to ensure proper sampling of the
source in the horizontal and vertical directions) and the
maximum number of vertical modes, $J$ (to ensure proper sampling
of the smallest wavelengths in each vertical mode).
The grid number in the depth direction is purposely chosen to be a power of
$2$ to allow the use of a fast Fourier transform for the split-operator
Fourier method.  Specifically, for the source frequencies
$25,75,150,250$ Hz, we chose $\Delta r = 0.01,0.01,0.005, 0.0025$ km
and $\Delta z = 9/N$ km where
$N=1024,2048,2048,2048$, respectively.  These values are
large enough to guarantee proper convergence of the
split-operator method for the propagation.

\section{IMPLEMENTATION OF INTERNAL WAVE SOUND SPEED MODEL}
\label{ap:soundspeed}

The efficient numerical scheme devised by Colosi and
Brown~\cite{colosibrown97} generates a random ensemble of internal wave
effects for the sound speed model, $\delta c_{iw}(z,r)/c_0$, through the
following equation:
\begin{eqnarray}
&&\frac{\delta c_{iw}(z,r)}{c_0}=
\frac{24.5}{g}\frac{2B}{\pi} N_0^2 \sqrt{\frac{E \Delta k_r}{M}}
\exp(-3z/2B) \hspace{0.1in} \\
&&\times\sum_{j=1}^{J_{max}}
\sum_{k_r=k_{min}}^{k_{max}} \sin(j\pi\xi(z))
\sqrt{\frac{I_{j,k_r}}{j^2+j_*^2}} \cos\left( \phi_{j,k_r}+k_r r\right) \nonumber
\end{eqnarray}
where
\begin{eqnarray}
k_j I_{j,k_r} = \frac{1}{\beta^2 +1}+
\frac{1}{2}\frac{\beta^2}{(\beta^2
+1)^{\frac{3}{2}}} \ln\left(\frac{\sqrt{\beta^2+1}+1}{\sqrt{\beta^2+1}-1}
\right)
\end{eqnarray}
A single random seed generates the random phases, $\phi_{j,k_r}\in[0,2\pi)$, 
for
each internal wave with vertical mode, $j$, and horizontal wavenumber, $k_r$.
These random phases give the ocean a different internal wave realization
for each random seed. All calculations in this paper were done with a single
realization of the internal wave field, but all results are similar for 
averages
over ensemble of internal wave fields as well.
Each internal wave in the superposition has the
statistics of the Garrett-Munk spectrum. The full Garrett-Munk energy of
$E= 6.3 \mbox{ x }  10^{-5}$ has been used in all calculations.
Our calculations are done for a latitude of $30^\circ$ so that
the inertial frequency is $f_i=1$ cycle per day.
The buoyancy profile is assumed to have the form $N(z)=N_0 e^{-z/B}$,
where $N_0=$ 1 cycle per 10 min is the buoyancy frequency at the surface.
We considered the depth of the ocean to be $H=5.0$ km,
even though we extended the propagation range to the region $[-2,7]$ km
for the reasons described in Appendix~\ref{ap:ssofft}.

The particular functional forms and constants used in this paper are as
used by Colosi and Brown.  
Some of these forms and constants have already been
identified in the body of the paper (i.e. near Eq.~(\ref{eq:gen_iw})),
while the others are listed here.
We took the gravitational acceleration
$g=9.81$ m/s$^2$, $M=(\pi j_* -1)/2 j_*^2$ and the principle mode
number $j_* = 3$.  We took $512$ horizontal internal wave numbers equally
spaced by $\Delta k_r$ for
$k_r\in 2\pi[0.01,1.0]$ cycles per km.
In the expression for $I_{j,k_r}$, we
took $k_j = f_i\pi j/N_0 B$ and the ratio $\beta =k_r/k_j$.


\end{document}